\documentclass[preprint,preprintnumbers,amsmath,amssymb]{revtex4}
\pdfoutput=1
\usepackage{epsfig}
\usepackage{graphicx}
\usepackage{dcolumn}
\usepackage{bm}
\usepackage{threeparttable}
\usepackage{xcolor}
\usepackage[utf8]{inputenc}
\usepackage{graphicx}
\usepackage{subfigure}

\def\beq{\begin{equation}}
\def\eeq{\end{equation}}
\def\eeqn{\end{equation}}
\newcommand\iden{\leavevmode\hbox{\small1\normalsize\kern-.33em1}}

\newcommand{\bea} {\begin{eqnarray}}
\newcommand{\eea} {\end{eqnarray}}

\def\tb {t_\beta}
\def\sb  {s_{\beta}}
\def\cb  {c_{\beta}}

\def\lam{\lambda}

\newcommand{\order}{{\cal O}}

\let\jnfont=\rm
\def\NPB#1 {{\jnfont Nucl.\ Phys.\ B }{\bf #1} }
\def\PLB#1 {{\jnfont Phys.\ Lett.\ B }{\bf #1} }
\def\EPJC#1 {{\jnfont Eur.\ Phys.\ Jour.\ C }{\bf #1} }
\def\PRD#1 {{\jnfont Phys.\ Rev.\ D }{\bf #1} }
\def\PRL#1 {{\jnfont Phys.\ Rev.\ Lett.\ }{\bf #1} }
\def\MPLA#1 {{\jnfont Mod.\ Phys.\ Lett.\ A }{\bf #1} }
\def\JPG#1 {{\jnfont J.\ Phys.\ G }{\bf #1} }
\def\CTP#1 {{\jnfont Commun.\ Theor.\ Phys.\ }{\bf #1} }
\def\JHEP#1 {{\jnfont JHEP \ }{\bf #1} }
\def\NPPS#1 {{\jnfont Nucl.\ Phys.\ Proc.\ Suppl.\ }{\bf #1} }
\def\CPC#1 {{\jnfont Comput.\ Phys.\ Commun.\ }{\bf #1} }
\def\CPL#1 {{\jnfont Chin.\ Phys.\ Lett. }{\bf #1} }
\def\APPB#1 {{\jnfont Acta\ Phys.\ Polon.\ B }{\bf #1} }

\def\lsim{\raise0.3ex\hbox{$<$\kern-0.75em\raise-1.1ex\hbox{$\sim$}}}
\def\gsim{\raise0.3ex\hbox{$>$\kern-0.75em\raise-1.1ex\hbox{$\sim$}}}
\def\PR#1 {{\jnfont Phys.\ Rept. }{\bf #1} }
\def\CHC#1 {{\jnfont Chin.\ Phys.\ C }{\bf #1} }
\def\NIMA#1 {{\jnfont Nucl.\ Instrum.\ Meth.\ A }{\bf #1} }
\def\JCAP#1 {{\jnfont JCAP \ }{\bf #1} }
\def\ASA#1 {{\jnfont Astron.\ Astrophys.\ A }{\bf #1} }  

\begin{document}

\title{\ \\[10mm] Spontaneous CP violation electroweak baryogenesis and gravitational wave through multi-step phase transitions }

\author{Songtao Liu, Lei Wang}
 \affiliation{Department of Physics, Yantai University, Yantai
264005, P. R. China
}


\begin{abstract}
In a singlet pseudoscalar extension of the two-Higgs-doublet model we discuss spontaneous CP violation electroweak baryogenesis via two different
patterns of phase transitions (PTs): (i) two-step PTs whose first-step and second-step are strongly first-order; (ii) three-step PTs whose first-step is
second-order and the second-step and third-step are strongly first-order. For the case of the two-step pattern, 
the first-step PT takes place at a high temperature, converting the origin phase into an
 electroweak symmetry broken phase and breaking the CP symmetry spontaneously. Thus, the baryon number is produced during the first-step PT.
At the second-step PT, the phase is converted into the observed vacuum at zero temperature, and
the CP-symmetry is restored. In both phases the sphaleron processes are sufficiently suppressed, which keep the baryon number unchanged.
For the case of the three-step PTs, the pseudoscalar field firstly acquires a nonzero VEV, and VEVs of other fields still remain zero during the first-step PT.
The following PTs and electroweak baryogenesis are similar to the case of the two-step PTs.
In addition, the gravitational wave spectra can have one or two
peaks through the two-step and the three-step PTs, and we discuss the detectability at the future gravitational wave detectors.
\end{abstract}

\maketitle

\section{Introduction}
The baryon asymmetry of the universe (BAU) is one of the longstanding questions of particle
physics and cosmology. The observed BAU from the Big Bang Nucleosynthesis
is given by \cite{pdg2020}
\beq
Y_B \equiv \rho_B/s = (8.2 - 9.2) \times 10^{-11},
\eeq  
 where $\rho_B$ is the baryon number density and $s$ is the entropy density.
Generating such an asymmetry dynamically
need satisfy the well-known Sakharov conditions: baryon number
violation, sufficient C and CP violation, and departure from thermal equilibrium \cite{Sakharov}.
A theoretically attractive mechanism is provided by the electroweak baryogenesis (EWBG) \cite{ewbg1,ewbg2}, which can be tested at current or future colliders because
it generally involves new physics around TeV.
In the EWBG scenario the baryon number is violated by sphaleron process at high temperatures, and the out-of-equilibrium environment is realized by a strong first-order electroweak
phase transition (SFOEWPT).
The SM contains the electroweak sphaleron process, but it fails to provide the out-of-equilibrium  and sufficient CP-violation. 
Therefore, a successful EWBG asks for an extension of the SM with additional sources of CP violation and extra particles coupling to the Higgs sector producing a SFOEWPT,
which can be realized in some typical extensions of the SM, such as the singlet extension of SM (see e.g. \cite{bgs-1,bgs-2,bgs-3,bgs-4,bgs-5,bgs-6,bgs-7,Huang:2018aja,Xie:2020wzn,cao,huang}) and the two-Higgs-doublet model (2HDM) (see e.g. \cite{bg2h-1,bg2h-2,bg2h-3,bg2h-4,bg2h-5,bg2h-6,bg2h-7,bg2h-8,bg2h-9,bg2h-10,bg2h-11,bg2h-12,Basler:2020nrq,bg2h-13,2111.13079,2207.00060}). 

The explicit breaking of CP may appear in the scalar
couplings or Yukawa couplings, which can be severely constrained by the non-observation of electric dipole moment (EDM) experiments \cite{edm-e}.
Several cancellation mechanisms are proposed to make the CP violation to be large enough to achieve the EWBG while satisfying the EDM data \cite{1411.6695,2004.03943,2111.13079,2207.00060}. 
On the other hand, a finite temperature spontaneous CP violation mechanism can naturally avoid the constraints of the EDM data, where the CP
symmetry is spontaneously broken at the high temperature and it is restored after the electroweak PT. 
 The spontaneous CP violation EWBG can be realized in the singlet complex scalar extension of the SM \cite{cao,huang}
and the singlet pseudoscalar extension of 2HDM \cite{Huber:2022ndk}, in which the the singlet field firstly acquires a nonzero vacuum expectation value (VEV) 
while the electroweak symmetry remains unbroken. Next, a SFOEWPT takes place through the vacuum decay between the singlet field direction and the doublet field direction in which
the net BAU is produced via the conventional EWBG mechanism.

In this paper, we discuss the spontaneous CP violation EWBG via two different
patterns of PTs in the singlet pseudoscalar extension of 2HDM: (i) two-step PTs whose first-step and second-step are strongly first-order; (ii) three-step PTs whose first-step is
second-order and the second-step and third-step are strongly first-order. 
In addition, the gravitational wave (GW) spectra could have one or two
peaks through the two-step and the three-step PTs \cite{2106.03439,2212.07756}, and we discuss the detectability at the future GW detectors, such as LISA
\cite{lisa}, Taiji \cite{taiji}, TianQin \cite{tianqin}, Big Bang Observer (BBO) \cite{bbodecigo}, DECi-hertz Interferometer
GW Observatory (DECIGO) \cite{bbodecigo} and Ultimate-DECIGO (UDECIGO) \cite{udecigo}.

Our work is organized as follows. In Sec. II we will give a brief introduction on the model. 
 In Sec. III and Sec. IV, we discuss the possibility of explaining the BAU and detecting the GW signal at
the future space-based detectors.
Finally, we give our conclusion in Sec. V.

\section{A singlet pseudoscalar extension of 2HDM}
A singlet pseudoscalar $S$ is introduced to the 2HDM, and the Higgs potential includes two parts: $\mathrm{V}
_{2HDM}$ and $\mathrm{V}_{S}$. They are respectively the pure potential of 2HDM and the potential containing the pseudoscalar $S$.
The $\mathrm{V}_{2HDM}$ with a softly broken discrete $Z_2$ symmetry is written as
\begin{eqnarray}
\label{V2HDM} \mathrm{V}_{2HDM} &=& m^2_{11} (\Phi_1^{\dagger} \Phi_1)
+ m^2_{22} (\Phi_2^{\dagger}
\Phi_2) + \left[m^2_{12} \Phi_1^{\dagger} \Phi_2 + \rm h.c.\right]\nonumber \\
&&+ \lambda_1  (\Phi_1^{\dagger} \Phi_1)^2 +
\lambda_2 (\Phi_2^{\dagger} \Phi_2)^2 + \lambda_3
(\Phi_1^{\dagger} \Phi_1)(\Phi_2^{\dagger} \Phi_2) + \lambda_4
(\Phi_1^{\dagger}
\Phi_2)(\Phi_2^{\dagger} \Phi_1) \nonumber \\
&&+ \left[\lambda_5 (\Phi_1^{\dagger} \Phi_2)^2 + \rm
h.c.\right].
\end{eqnarray}
 The $\Phi_1$ and $\Phi_2$ are complex Higgs doublets with hypercharge $Y = 1$:
\begin{equation}
\Phi_1=\left(\begin{array}{c} \phi_1^+ \\
\frac{1}{\sqrt{2}}\,(v_1+\phi_1^0+i\eta)
\end{array}\right)\,, \ \ \
\Phi_2=\left(\begin{array}{c} \phi_2^+ \\
\frac{1}{\sqrt{2}}\,(v_2+\phi_2^0+ih_3)
\end{array}\right).
\end{equation}
Where $v_1$ and $v_2$ are the electroweak VEVs with $v^2 = v^2_1 + v^2_2 = (246~\rm GeV)^2$, and the ratio of the two VEVs is defined
as $\tan\beta=v_2 /v_1$. 

The $\mathrm{V}_{S}$ containing the singlet pseudoscalar $S$ is given by
\beq\label{V2hs}
\mathrm{V}_{S}=\frac{1}{2}m^2_{0}S^2+\frac{\kappa_{S}}{24}S^4 + \left[i\mu S \Phi_2^{\dagger} \Phi_1 +h.c.\right] + \frac{\kappa_1}{2}S^2\Phi_1^{\dagger} \Phi_1+ \frac{\kappa_2}{2}S^2\Phi_2^{\dagger}\Phi_2.
\eeq
Here we assume that all coupling coefficients and mass terms are real, and the pseudoscalar $S$ does not develop a VEV at zero temperature.
As a result, the Higgs potential sector is CP-conserved at zero temperature.

The potential minimization conditions require
\beq
\begin{split}
&\quad m_{11}^2 = m_{12}^2 \tb - \frac{1}{2} v^2 \left( \lambda_1 \cb^2 + \lambda_{345}\sb^2 \right)\,,\\
& \quad m_{22}^2 =  m_{12}^2 / \tb - \frac{1}{2} v^2 \left( \lambda_2 \sb^2 + \lambda_{345}\cb^2 \right)\,,\\
& \quad m_{0}^2 + \frac{\kappa_1}{2} v^2 \cb^2 +  \frac{\kappa_2}{2} v^2 \sb^2 > 0\,,
\end{split}
\label{min_cond}
\eeq
where the shorthand notations  $\tb\equiv \tan\beta$, $\sb\equiv \sin\beta$, $\cb \equiv \cos\beta$,
and $\lambda_{345} = \lambda_3+\lambda_4+\lambda_5$.

After spontaneous electroweak symmetry breaking, the remaining physical states are two neutral CP-even states $h$ and $H$,
 two neutral pseudoscalars $A$ and $X$, and a pair of charged scalars $H^{\pm}$. 
The sources of mass eigenstates $h$, $H$ and $H^{\pm}$ and their masses are the same as those of the pure 2HDM. 
In addition to $A$ and $X$, the Goldstone boson $G$ is also one mass eigenstate of pseudoscalar, and they are from the mixing
of $\eta$, $h_3$ and $S$ with two mixing angles $\theta$ and $\beta$. 
The parameters $m_0^2$ and $\mu$ are determined by
\beq
\begin{split}
&\quad \mu = \frac{m_{X}^2-m_A^2}{v}s_\theta c_\theta\,,\\
& \quad m_0^2 = m_A^2 s_\theta^2+ m_X^2 c_\theta^2  -\frac{\kappa_1}{2} v^2 \cb^2 - \frac{\kappa_2}{2} v^2 \sb^2\,,
\end{split}
\label{eq:mum0}
\eeq
where $s_\theta\equiv \sin\theta$ and $c_\theta \equiv \cos\theta$.

The general Yukawa interactions are written as
 \bea
- {\cal L} &=&Y_{u2}\,\overline{Q}_L \, \tilde{{ \Phi}}_2 \,u_R
+\,Y_{d2}\,
\overline{Q}_L\,{\Phi}_2 \, d_R\, + \, Y_{\ell 2}\,\overline{L}_L \, {\Phi}_2\,e_R \,\nonumber\\
&+&Y_{u1}\,\overline{Q}_L \, \tilde{{ \Phi}}_1 \,u_R
+\,Y_{d1}\,
\overline{Q}_L\,{\Phi}_1 \, d_R\, + \, Y_{\ell 1}\,\overline{L}_L \, {\Phi}_1\,e_R+\, \mbox{h.c.}\,,
\eea where
$Q_L^T=(u_L\,,d_L)$, $L_L^T=(\nu_L\,,l_L)$,
$\widetilde\Phi_{1,2}=i\tau_2 \Phi_{1,2}^*$, and $Y_{u1,2}$,
$Y_{d1,2}$ and $Y_{\ell 1,2}$ are $3 \times 3$ matrices in family
space. In order to avoid the tree-level flavour changing neutral current, we take the Yukawa interactions to be aligned \cite{aligned2h},
 \bea
 &&(Y_{u1})_{ii}=\frac{\sqrt{2}m_{ui}}{v}(c_\beta-s_\beta \kappa_u),~~~~~(Y_{u2})_{ii}=\frac{\sqrt{2}m_{ui}}{v}(s_\beta+c_\beta \kappa_u),\nonumber\\
&&(Y_{\ell 1})_{ii}=\frac{\sqrt{2}m_{\ell i}}{v}(c_\beta-s_\beta \kappa_\ell),~~~~~(Y_{\ell 2})_{ii}=\frac{\sqrt{2}m_{\ell i}}{v}(s_\beta+c_\beta \kappa_\ell),\nonumber\\
&&(X_{d1})_{ii}=\frac{\sqrt{2}m_{di}}{v}(c_\beta-s_\beta \kappa_d),~~~~~(X_{d2})_{ii}=\frac{\sqrt{2}m_{di}}{v}(s_\beta+c_\beta \kappa_d).
\eea
Where all the off-diagonal elements are zero. $i=1,2,3$ is the index of generation and $X_{d1,2}=V_{CKM}^\dagger Y_{d1,2} V_{CKM}$.

\section{Electroweak phase transition and baryogenesis}
\subsection{Relevant theoretical and experimental constraints}
Before discussing the electroweak PT and EWBG, we first introduce relevant theoretical and experimental constraints.
We identify the lightest CP even Higgs boson $h$ as the observed 125 GeV state, and take $\sin(\beta-\alpha)=1$ in order to avoid
the constraints of the 125 GeV Higgs signal data, for which the tree-level couplings of $h$ to the SM particles are the same
to the SM. The $h$ is assumed to have no exotic decay mode. In addition, we assume $\kappa_u$, $\kappa_d$ and $\kappa_\ell$ to be small enough so that the 
extra Higges ($H$, $H^\pm$, $A$, $X$) can satisfy the exclusion limits of searches for additional Higgs bosons at the collider and 
the constraints of flavor observables. Also the other effects induced by the three parameters are ignored in the following discussions.

The scalar potential of the model includes the potential of 2HDM and the potential involved the singlet field $S$, which are
constrained by the vacuum stability, perturbativity, and tree-level unitarity.
There are detailed discussions in Refs. \cite{2hisos-4,2hisos-6}, and we employ the formulas in \cite{2hisos-4,2hisos-6} to
implement the theoretical constraints.
 The model can give additional corrections to the oblique parameters ($S$, $T$, $U$) via 
the self-energy diagrams exchanging extra Higgs fields ($H$, $H^\pm$, $A$, $X$). For $\sin(\beta-\alpha)=1$, the expressions of $S$, $T$ and $U$ in the 
this model are approximately given as \cite{stu1,stu2}
\bea
S&=&\frac{1}{\pi m_Z^2}
\left[  c_\theta^2 F_S(m_Z^2,m_H^2,m_A^2) + s_\theta^2 F_S(m_Z^2,m_H^2,m_X^2)
-F_S(m_Z^2,m_{H^{\pm}}^2,m_{H^{\pm}}^2) \right], \nonumber \\
T&=&\frac{1}{16\pi m_W^2 s_W^2} \left[ - c_\theta^2 F_T(m_H^2,m_A^2) - s_\theta^2 F_T(m_H^2,m_X^2)
+ F_T(m_{H^{\pm}}^2,m_H^2) \right. \nonumber \\
&&+ \left. c_\theta^2 F_T(m_{H^{\pm}}^2,m_A^2) + s_\theta^2 F_T(m_{H^{\pm}}^2,m_X^2)\right], \nonumber\\
U&=&\frac{1}{\pi m_W^2} \left[ F_S(m_W^2,m_{H^{\pm}}^2,m_H^2) -2 F_S(m_W^2,m_{H^{\pm}}^2,m_{H^{\pm}}^2) \right.\nonumber\\
&& \left.+c_\theta^2 F_S(m_W^2,m_{H^{\pm}}^2,m_A^2) + s_\theta^2 F_S(m_W^2,m_{H^{\pm}}^2,m_X^2) \right]\nonumber\\
&&-\frac{1}{\pi m_Z^2} \left[  c_\theta^2 F_S(m_Z^2,m_H^2,m_A^2) + s_\theta^2 F_S(m_Z^2,m_H^2,m_X^2) \right.\nonumber\\
&&\left.- F_S(m_Z^2,m_{H^{\pm}}^2,m_{H^{\pm}}^2) \right],
\eea
where
\beq
F_T(a,b)=\frac{1}{2}(a+b)-\frac{ab}{a-b}\log(\frac{a}{b}),~~F_S(a,b,c)=B_{22}(a,b,c)-B_{22}(0,b,c),
\eeq
with
\bea
&&B_{22}(a,b,c)=\frac{1}{4}\left[b+c-\frac{1}{3}a\right] - \frac{1}{2}\int^1_0 dx~X\log(X-i\epsilon),\nonumber\\
&&
X=bx+c(1-x)-ax(1-x).
\eea
Taking the recent fit results of Ref. \cite{pdg2020}, we use the following 
values of $S$, $T$, $U$,
\beq
S=-0.01\pm 0.10,~~  T=0.03\pm 0.12,~~ U=0.02 \pm 0.11, 
\eeq
with the correlation coefficients 
\beq
\rho_{ST} = 0.92,~~  \rho_{SU} = -0.80,~~  \rho_{TU} = -0.93.
\eeq

\subsection{Electroweak phase transition and bubble profiles}
To analyze the electroweak PT, one needs the effective potential of the model at the finite temperature.
We parameterize the neutral components of the two Higgs doublets,
\beq
\frac{1}{\sqrt{2}}(h_1+i\eta)=\frac{1}{\sqrt{2}}A_1 e^{i\varphi_1},~~~~~\frac{1}{\sqrt{2}}(h_2+ih_3)=\frac{1}{\sqrt{2}}A_2 e^{i\varphi_2}.
\eeq
From Eq. (\ref{V2HDM}) and Eq. (\ref{V2hs}), one finds that the effective potential only depends on the relative phase $\varphi_2-\varphi_1$.
Thus, we choose to rotate $\varphi_1$ to 0, and take $h_1$, $h_2$, $h_3$ and $S$ as the field configurations. 
The complete effective potential at finite temperature includes the tree level potential, the
Coleman-Weinberg term \cite{vcw}, the finite temperature corrections \cite{vloop} and the resummed daisy corrections \cite{vring1,vring2}, which is gauge-dependent \cite{vgauge1,vgauge2}. 
Here we take a gauge invariant approximation, which keeps only the thermal mass terms in the high-temperature
expansion in addition to the tree level potential. Then the effective potential is written as
\begin{align}
V_{eff} (h_1,h_2,a_2,S_1,T)= & \frac{1}{2} (m_{11}^2 + \Pi_{h_1}) h_1^2 + \frac{1}{2} (m_{22}^2+ \Pi_{h_2}) h_2^2 + \frac{1}{2} (m_{22}^2 + \Pi_{h_3}) h_3^2 - m_{12}^2 h_1 h_2\nonumber\\
&+ \frac{\lambda_1}{8} h_1^4 +  \frac{\lambda_2}{8} (h_2^4 + h_3^4) + \frac{\bar{\lambda}_{345}}{4} h_1^2 h_3^2+ \frac{\lambda_{345}}{4} h_1^2 h_2^2+ \frac{\lambda_{2}}{4} h_2^2 h_3^2\nonumber\\
&+ \frac{1}{2} (m_0^2+\Pi_{S}) S^2 + \mu S h_1 h_3 + \frac{\kappa_1}{4} S^2 h_1^2 + \frac{\kappa_2}{4} S^2 (h_2^2 + h_3^2) + \frac{\kappa_S}{24} S^4,
\label{veff0}
\end{align}
with
\begin{align}
\Pi_{h_1} &= \left[{9g^2\over 2} + {3g'^2\over 2} + 6\lam_{1} +4\lam_{3} +2\lam_4 + \kappa_{1} + 6y_t^2 c_\beta^2\right] {T^2 \over 24},\nonumber\\
\Pi_{h_2} &= \left[{9g^2\over 2} + {3g'^2\over 2} + 6\lam_{2} +4\lam_{3} +2\lam_4 + \kappa_{2} + {6y_t^2 s_\beta^2}\right] {T^2 \over 24},\nonumber\\
\Pi_{h_3} &=\Pi_{h_2},\nonumber\\
\Pi_{S} &= \left[4\kappa_{1} +4\kappa_{2} +\kappa_S \right] {T^2 \over 24},
\end{align}
where $\bar{\lambda}_{345}=\lambda_3+\lambda_4-\lambda_5$ and  $y_t={\sqrt{2} m_t \over v}$.

In a first-order PT, bubbles nucleate and expand, converting
the high-temperature phase into the low-temperature one.
The probability of tunneling at the temperature $T$ per unit time per unit volume is is given by \cite{bubble-0,bubble-1,bubble-2}
\begin{eqnarray}
\Gamma \ \approx \ A(T)e^{-S_3/T},
\end{eqnarray}
where $A(T)\sim T^4$ is a prefactor and $S_3$ is a three-dimensional Euclidian action.

The Euclidian action is calculated with $O(3)$ symmetric solutions for the configurations $h_1$, $h_2$, $h_3$, and $S$, which are determined by differential equations~\cite{vaccum-eq}
\begin{equation}
\label{eq: bubble_equation}
\frac{ \mathrm{d}^2 \varphi_i }{ \mathrm{d} r^2 } + \frac{ 2 }{ r } \frac{ \mathrm{d} \varphi_i }{ \mathrm{d} r } = \frac{ \partial V_{eff} }{ \partial \varphi_i }, \quad ( i=1,2,3,4), 
\end{equation}
with the boundary conditions  $\mathrm{d} \varphi_i  / \mathrm{d} r |_{r=0} = 0$ and $\varphi_i (r = \infty) = \varphi_{if}$ with $\varphi_{if}$ being VEV of phase outside bubble.
Here $\varphi_{i=1,2,3,4}$ denote $h_1$, $h_2$, $h_3$, and $S$, and $r$ is the spatial radial coordinate. 
The solutions are used to determine the value of $S_3$,
\begin{equation}
S_3 = 4 \pi \int_0^\infty \mathrm{d}r \, 
r^2 \left [
	\sum_{i=1}^4 \frac{ 1 }{ 2 } 
		\left( 
		\frac{ \mathrm{d} \varphi_i }{ \mathrm{d} r }
		\right)^2
	+ V_{eff}
\right ]. 
\end{equation}
 At the nucleation temperature $T_n$, the thermal tunneling probability for bubble nucleation per horizon volume and per horizon time is 
of order one, and the conventional condition is $\frac{S_3(T)}{T}|_{T = T_n}\approx 140$. 

The dynamics of the electroweak PT are characterized by two key parameters $\beta$ and $\alpha$. $\beta$ characterizes roughly the inverse time duration of the strong first-order PT,
\begin{eqnarray}
\frac{\beta}{H_n}=T\frac{d (S_3(T)/T)}{d T}|_{T=T_n}\; ,
\end{eqnarray}
where $H_n$ is the Hubble parameter at the nucleation temperature $T_n$.
$\alpha$ is defined as the vacuum energy released from the phase transition normalized by the total radiation energy
density $\rho_R$ at $T_n$,
      \begin{eqnarray}
      \alpha=\frac{\Delta\rho}{\rho_R}=\frac{\Delta\rho}{\pi^2 g_{\ast} T_n^4/30}\;,
      \end{eqnarray}
where $g_{\ast}$ is the effective number of relativistic degrees of freedom.

During the first SFOEWPT, the doublet fields develop nonzero VEVs, namely yielding a transition from (0,~0,~0) to ($<h_1>$, $<h_2>$, $<h_3>$). 
The CP violation comes directly from the spatial evolution of $<h_3>$, which renders the top quark mass
a complex-valued function of the spatial coordinate across the bubble wall. The electroweak sphaleron processes \cite{sphaleron-1,sphaleron-2,sphaleron-3} can bias the
CP asymmetry produced around the bubble wall into the baryon asymmetry. The condition that guarantees the produced baryon asymmetry inside the bubbles
of broken phase is not washed out by the electroweak sphalerons, leads to a bound on the PT strength \cite{pt-stren}
\beq
\frac{\xi_{n1}}{T_{n1}} > 1.0,
\eeq 
where $\xi_{n1}=\sqrt{<h_1>^2+ <h_2>^2 + <h_3>^2}$ and $T_{n1}$ is the nucleation temperature for the first SFOEWPT.

As temperature drops down to the $T_{n2}$, the second SFOEWPT takes place, in which the phase is converted into the observed vacuum at zero temperature, and the CP symmetry is restored.
During the second SFOEWPT, the sphaleron processes are sufficiently suppressed in both phases, which keep the baryon number unchanged.


\begin{table}
\begin{footnotesize}
\begin{tabular}{| c | c | c | c |c | c | c | c |c | c | c | c |c | c | c | c |}
\hline
 &$\tan\beta$ & $m_{12}^2$(GeV$)^2$& $m_H$(GeV) &$m_A$(GeV)& $m_{H^\pm}$(GeV) & $m_X$(GeV) & $s_\theta$ & $\kappa_1$& $\kappa_2$ &$\kappa_s$\\
\hline
$\textbf{BP1}$   & ~~0.867~~& ~~4628.4~~ &~~ 463.3~~& ~~124.6~~ &~~ 478.2~~& ~~539.2~~ &~~ -0.372~~& ~~9.294~~ &~~7.176~~& ~~0.881~~  \\
 \hline
$\textbf{BP2}$   & ~~1.084~~& ~~2864.9~~ &~~ 481.6~~& ~~161.5~~ &~~ 494.4~~& ~~412.6~~ &~~ -0.325~~& ~~7.198~~ &~~4.361~~& ~~10.677~~  \\
 \hline
\end{tabular}
\begin{tabular}{ c |  c }
~~~~~~~~~~The ~first~ SFOEWPT~~~~~~&~~~~~~~~~~~The~ second~SFOEWPT~~~~~~~~\\  
\end{tabular}

\begin{tabular}{| c | c | c | c |c | c | c | c |c | c | c | c |c | c | c | c |}
\hline
 & ~~$T_{n1}$(GeV)~~ & ~~$\xi_{n1}/T_{n1}$~~ & ~~$T_{n2}$(GeV)~~ & ~~$\xi_{n2}/T_{n2}$~~& ~~$\xi'_{n2}/T_{n2}$~ \\
\hline
$\textbf{BP1}$ & ~~ 71.86 ~~ & ~~1.24~~ & ~~34.82~~ &~~6.11~~&~~4.72~~\\
\hline
$\textbf{BP2}$ & ~~ 62.48 ~~ & ~~1.20~~ & ~~58.11~~ &~~2.39~~&~~1.66~~\\
\hline
\end{tabular}

\end{footnotesize}
\caption{Input and output parameters for the BP1 and BP2 with $m_h$=125 GeV and $\sin(\beta-\alpha)=1$.
Here $\xi_{n2}$ and $\xi'_{n2}$ denote $\sqrt{<h_1>^2+<h_2>^2+<h_3>^2}$ of phases at the interior and exterior of the bubble from the second SFOEWPT. The first SFOEWPT
is from the first-step of two-step PTs and the second-step of three-step PTs, and the second SFOEWPT
is from the second-step of two-step PTs and the third-step of three-step PTs.}
\label{tabgrav}
\end{table}

\begin{figure}[tb]
 \epsfig{file=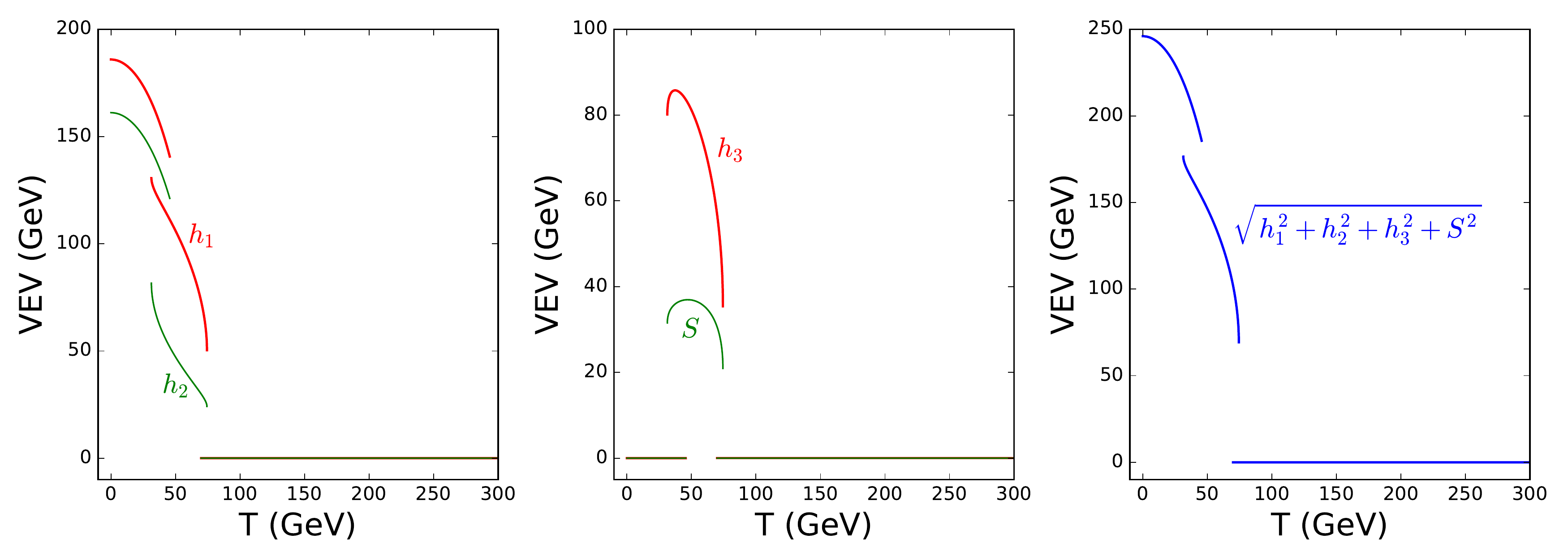,height=6.0cm}
\vspace{-1.0cm} \caption{Phase histories for the BP1.} \label{figphase}
\end{figure}

\begin{figure}[tb]
 \epsfig{file=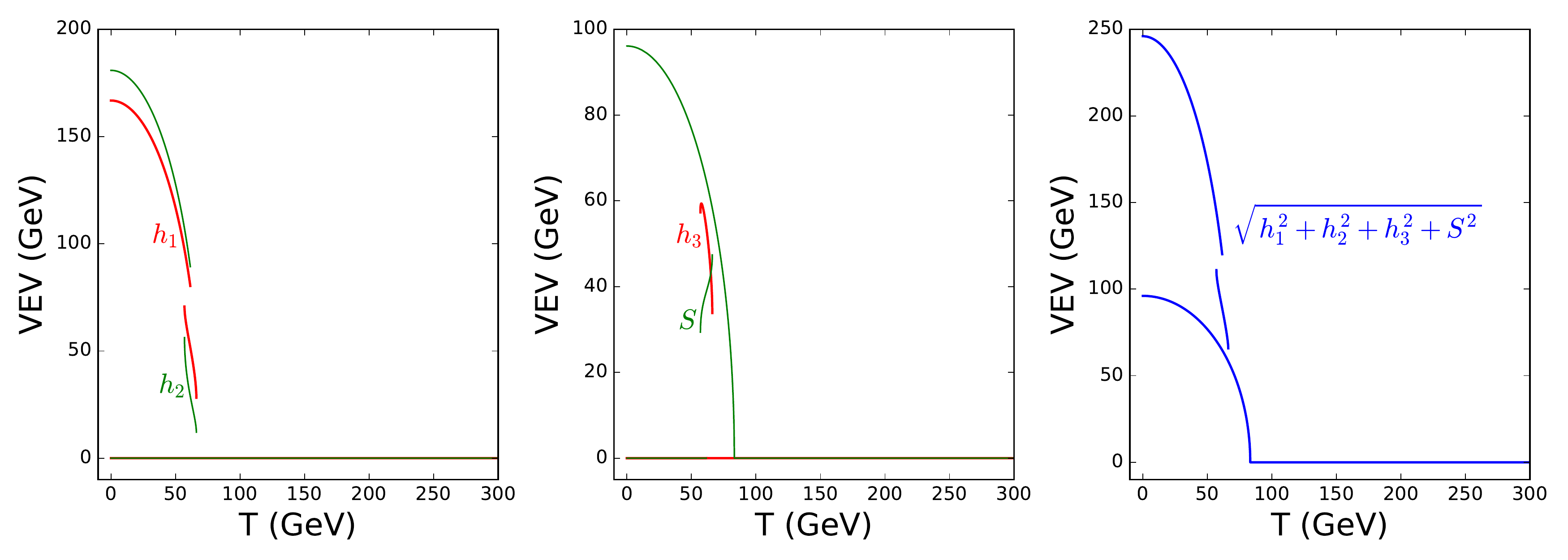,height=6.0cm}
\vspace{-1.0cm} \caption{Phase histories for the BP2.} \label{figphase2}
\end{figure}

In our calculations, we require that the potential has a global minimum at the point of ($<h_1>=v_1$, $<h_2>=v_2$, $<h_3>$=0, $<S>=0$) at zero temperature,
which is numerically calculated. Considering the theoretical and experimental constraints discussed above, 
we take benchmark point 1 (BP1) and benchmark point 2 (BP2) to provide detailed discussions on the physical processes for the two-step PTs and three-step PTs, respectively,
which are shown in Table. \ref{tabgrav}. The oblique parameters favor $m_H$ and $m_{H^\pm}$ to have a small mass splitting.
The pseudoscalar with a mass of 124.6 GeV for the BP1 is allowed by the signal data of the observed 125 GeV Higgs at the LHC since its couplings to $WW$ and $ZZ$ are absent,
and the couplings to fermions are taken to be negligibly small.
The phase histories for the BP1 and BP2 are respectively exhibited in Fig. \ref{figphase} and Fig. \ref{figphase2} on field configurations versus temperature plane.
The numerical package 
CosmoTransitions \cite{cosmopt} is used to analyze the PTs.
As the Universe cools, there appear three different phases and they follow two-step PTs for the BP1.
At very high temperatures, because of the contributions of the thermal mass terms, the minimum of the potential is at the origin and the
electroweak symmetry is restored, namely  ($<h_1>$, $<h_2>$, $<h_3>$, $<S>$) = (0,~0,~0,~0) GeV. When the temperature decreases to 71.86 GeV, 
the system tunnels to the second phase at (67.2, 28.1, 51.6, 28.5) GeV
 via the first SFOEWPT. Because $h_3$ and $S$ acquire nonzero VEVs, the CP symmetry is broken spontaneously. As the temperature decreases, the system evolves
along the second phase until T = 34.82 GeV and ($<h_1>$, $<h_2>$, $<h_3>$, $<S>$) = (123.5, 67.2, 85.4, 35.1) GeV. Then it 
tunnels to the third phase at (161.1, 139.2, 0, 0) GeV and the CP symmetry is restored via the second SFOEWPT. Next, the system evolves
along the third phase and ultimately ends in the observed vacuum at T = 0 GeV.

Fig. \ref{figphase2} shows that the universe undergoes three-step PTs for the BP2.
At T = 83 GeV, the $S$ field acquires a nonzero VEV, and the VEVS of $h_1$, $h_2$, and $h_3$ still remain zero via a second-order PT. When the temperature decreases to 62.48 GeV, 
the system tunnels to a new phase at (48.3, 25.8, 51.5, 39.8) GeV  
 via the first SFOEWPT. As the temperature decreases, the system evolves
along the phase until T = 58.11 GeV and ($<h_1>$, $<h_2>$, $<h_3>$, $<S>$) = (63.4, 42.7, 59.2, 34.3) GeV. Then it 
tunnels to the final phase at (93.4, 103.1, 0, 0) GeV via the second SFOEWPT, and ultimately ends in the observed vacuum at T = 0 GeV.

\begin{figure}[tb]
 \epsfig{file=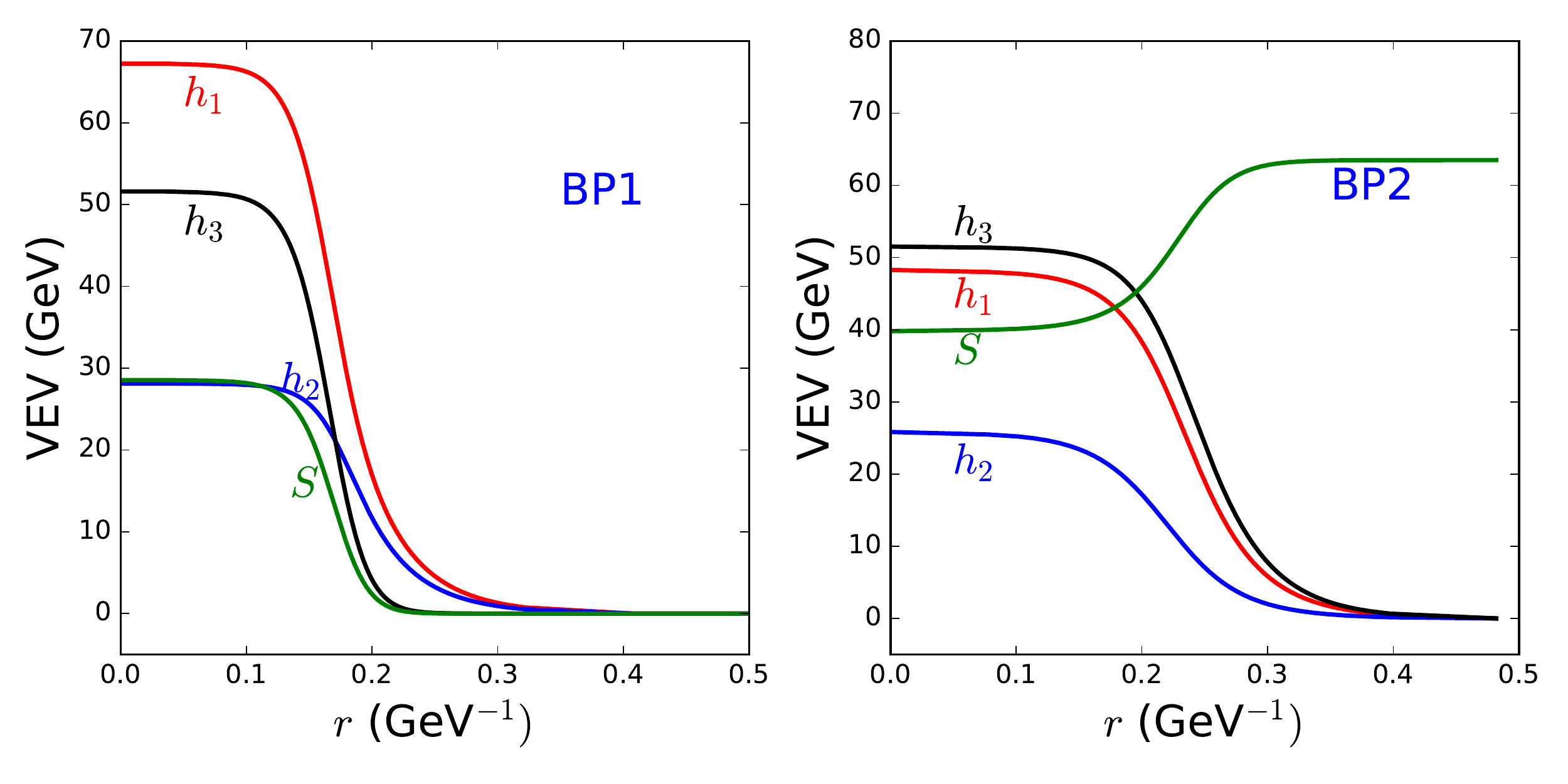,height=7.0cm}
\vspace{-0.3cm} \caption{The radial nucleation bubble wall VEV profiles of the first SFOEWPT for the BP1 and the BP2. Here $r = 0$ is the centre of the bubble.} \label{figprofile}
\end{figure}

The bubble wall VEV profiles are determined by the solutions of the bounce equations in Eq. (\ref{eq: bubble_equation}), which are approximately obtained by FindBounce \cite{findbounce}.  
The baryon number is produced during the first SFOEWPT, and the relevant calculation depends on the bubble wall profiles. Therefore, in Fig. \ref{figprofile} we show the wall profiles 
of the first SFOEWPT for the BP1 and the BP2.

\subsection{Transport equations and baryon asymmetry}
We take the WKB method to discuss the CP-violating source terms and chemical potentials transport equations of particle species in the wall frame with 
a radial coordinate $z$ \cite{bg2h-3,0006119,0604159}.
The bubble wall is located at $z = 0$, with $z < 0$ and $z > 0$ pointing toward the interior and exterior of the bubble.
In the model, the top quark plays the most important role in generating the BAU during the first SFOEWPT.
It acquires a complex mass as a function of $z$ when passing through the bubble wall whose profiles depend on the coordinate $z$.
The mass of top quark is given as
\bea
m_t(z)&=&\frac{y_t}{\sqrt{2}}e^{i\varphi_Z(z)}(c_\beta h_1(z)+ s_\beta \sqrt{h^2_2(z)+h^2_3(z)}e^{i\varphi_2(z)}),\nonumber\\
&=&\frac{y_t}{\sqrt{2}}\sqrt{(c_\beta h_1(z)+s_\beta h_2(z))^2+s_\beta^2 h^2_3(z)}~ e^{i\theta_t},
\eea
with
\bea
&&\varphi_2(z)=\arctan\frac{h_3(z)}{h_2(z)},~~\theta_t=\varphi_Z(z)+\arctan\frac{s_\beta h_3(z)}{c_\beta h_1(z)+s_\beta h_2(z)},\nonumber\\
&&\partial_z\varphi_Z(z)=-\frac{h^2_2(z)+h^2_3(z)}{h^2_1(z)+h^2_2(z)+h^2_3(z)}\partial_z\varphi_2(z).
\eea 
 The addition phase $\varphi_Z(z)$ 
is from a local axial transformation of top quark which removes the CP-violating force induced by the nonvanishing $Z_\mu$ field in the case of $\varphi_1=0$ \cite{bg2h-4}.

The transport equations are derived for the top quark with a complex mass term,
and include effects of the strong sphaleron process ($\Gamma_{ss}$) \cite{bg2h-3,9311367}, W-scattering ($\Gamma_W$) \cite{bg2h-3,9506477}, the top Yukawa interaction ($\Gamma_y$) \cite{bg2h-3,9506477},
 the top helicity flips ($\Gamma_M$) \cite{bg2h-3,9506477}, and the Higgs number violation ($\Gamma_h$) \cite{bg2h-3,9506477}. The transport equations are given by
	\begin{align}
		0 =   & 3 v_W K_{1,t} \left( \partial_z \mu_{t,2} \right) + 3v_W K_{2,t} \left( \partial_z m_t^2 \right) \mu_{t,2} + 3 \left( \partial_z u_{t,2} \right) \notag
		\\ &- 3\Gamma_y \left(\mu_{t,2} + \mu_{t^c,2} + \mu_{h,2} \right) - 6\Gamma_M \left( \mu_{t,2} + \mu_{t^c,2} \right) - 3\Gamma_W \left( \mu_{t,2} - \mu_{b,2} \right) \notag
		\\ &- 3\Gamma_{ss} \left[ \left(1+9 K_{1,t} \right) \mu_{t,2} + \left(1+9 K_{1,b} \right) \mu_{b,2} + \left(1-9 K_{1,t} \right) \mu_{t^c,2} \right] \notag \,,\\
                0=    & 3 v_W K_{1,t} \left( \partial_z \mu_{t^c,2} \right)  + 3v_W K_{2,t} \left( \partial_z m_t^2 \right)  \mu_{t^c,2} + 3 \left( \partial_z u_{t^c,2} \right) \notag   \\
		      & - 3\Gamma_y \left(\mu_{t,2} + \mu_{b,2} + 2\mu_{t^c,2} + 2\mu_{h,2} \right) - 6\Gamma_M \left( \mu_{t,2} + \mu_{t^c,2} \right) \notag    \\
		      & - 3\Gamma_{ss} \left[ \left( 1+9 K_{1,t}\right) \mu_{t,2} + \left(1+9K_{1,b}\right) \mu_{b,2} + \left(1-9K_{1,t}\right) \mu_{t^c,2} \right]  \notag \,,          \\
		0 =   & 3v_W K_{1,b} \left(\partial_z \mu_{b,2}\right) + 3 \left(\partial_z u_{b,2} \right) - 3\Gamma_y \left( \mu_{b,2} + \mu_{t^c,2} + \mu_{h,2} \right) - 3\Gamma_W \left( \mu_{b,2} - \mu_{t,2} \right) \notag \,,   \\
		      & - 3\Gamma_{ss} \left[ \left( 1 + 9K_{1,t}\right) \mu_{t,2} + (1+9K_{1,b}) \mu_{b,2} + (1-9K_{1,t}) \mu_{t^c,2} \right] \notag \,, \\
		0 =   & 4v_W K_{1,h} \left( \partial_z \mu_{h,2}\right) +
		4\left( \partial_z u_{h,2}\right) - 3\Gamma_y \left(
		\mu_{t,2} + \mu_{b,2} + 2\mu_{t^c,2} + 2\mu_{h,2} \right) -
		4\Gamma_h
		\mu_{h,2}  \notag\,,\\
		S_t = & -3K_{4,t} \left( \partial_z \mu_{t,2}\right) + 3v_W \tilde{K}_{5,t} \left( \partial_z u_{t,2}\right) + 3v_W \tilde{K}_{6,t} \left( \partial_z m_t^2 \right) u_{t,2} + 3\Gamma_t^\mathrm{tot} u_{t,2} \notag \,,       \\
		0 =   & -3K_{4,b} \left( \partial_z \mu_{b,2} \right) + 3v_W \tilde{K}_{5,b} \left(\partial_z u_{b,2}\right) + 3\Gamma_b^\mathrm{tot} u_{b,2}  \notag  \,,   \\
		S_t = & -3K_{4,t} \left( \partial_z \mu_{t^c,2}\right) + 3v_W \tilde{K}_{5,t} \left( \partial_z u_{t^c,2}\right) + 3v_W \tilde{K}_{6,t} \left( \partial_z m_t^2\right) u_{t^c,2} + 3\Gamma_t^\mathrm{tot} u_{t^c,2} \notag \,, \\
		0 =   & -4K_{4,h} \left( \partial_z \mu_{h,2} \right) + 4v_W \tilde{K}_{5,h} \left( \partial_z u_{h,2} \right) + 4\Gamma_h^\mathrm{tot} u_{h,2}  \,,
\label{TransportEquations}
\end{align}

The $\mu_{i,2}$ and $u_{i,2}$ are the second-order CP-odd chemical potential and the
plasma velocity of the particle $i=t,~t^c,~b,~h$, respectively. The source term $S_t$ is defined as
\beq \label{Eq:TransportEquations:Source}
	S_t = -v_W K_{8,t} \partial_z \left( m_t^2 \partial_z \theta_t \right) + v_W K_{9,t} \left( \partial_z \theta_t \right) m_t^2  \left( \partial_z m_t^2\right).
\eeq
The functions $K_{a,i}$ and $\tilde{K}_{a,i}$ ($a=1-9$) are defined in Ref. \cite{0604159}, and
the $\Gamma_{i}^{\mathrm{tot}}$ denotes the total reaction rate of the particle $i$ \cite{bg2h-3,0604159}.
We treat the wall velocity $v_W$ as an input parameter and take $v_W=0.1$. 

The WKB method of calculating the source terms and transport equations  is valid for $L_W T_n \gg 1$ with $L_W$ being the width of bubble wall. 
We evaluate $L_W$ by fitting the profile of the VEV with the hyperbolic tangent function,
\beq
\sqrt{h^2_1(z)+h^2_2(z)+h^2_3(z)}=\frac{v_{n}}{2}\left(1-\tanh\frac{z}{L_W}\right).
\eeq
Using this approach one obtains $L_W T_{n1} \simeq 3.13$ and $L_W T_{n1} \simeq 3.68$ for the bubble wall of the first SFOEWPT of the BP1 and the BP2, respectively.   

\begin{figure}[tb]
 \epsfig{file=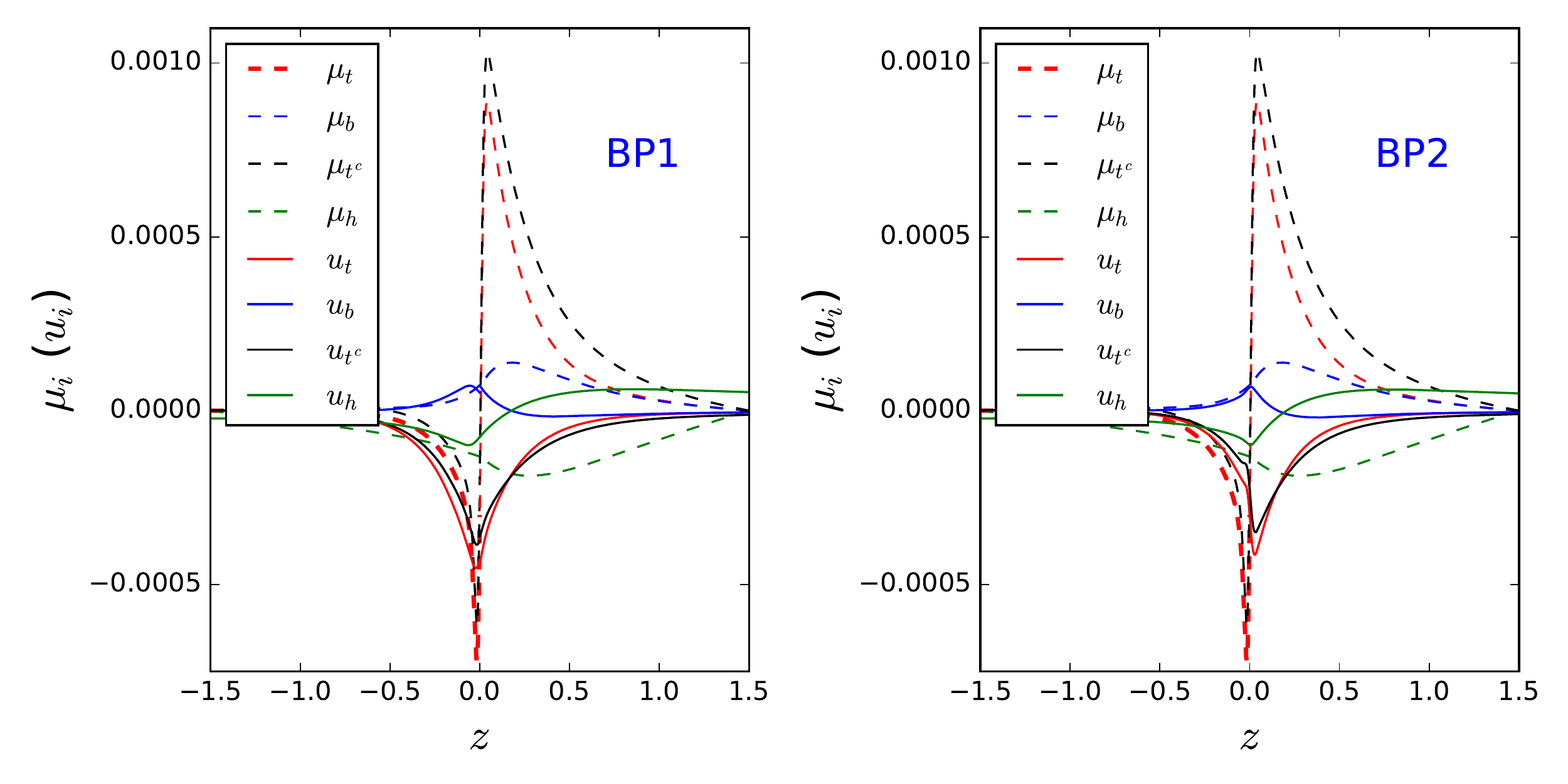,width=14cm,height=6.0cm}
\vspace{-0.8cm} \caption{For the BP1 and the BP2, the solutions to the transport equations for $\mu_i$ and $u_i$ as functions of the coordinate $z$ transverse to the bubble wall.} \label{figmu}
\end{figure}

One solves the transport equations with the boundary conditions $\mu_i$ $(z=\pm \infty) =0$ ($i=t,~t^c,~b,~h$), and obtains the chemical potentials $\mu_i$ of
each particle species. Using local baryon number conservation the chemical potential of the left-handed quarks is
then given by \cite{0006119}
\beq 
	\mu_{B_L} = \frac{1}{2} \left(1+4K_{1,t}\right) \mu_{t,2} + \frac{1}{2} \left(1+4K_{1,b}\right) \mu_{b,2} - 2K_{1,t} \mu_{t^c,2} \,.
\eeq
Next, the weak sphalerons convert the left-handed quark number into a baryon asymmetry, which can be calculated with 
\begin{equation}
	Y_B = \frac{405 \Gamma_{ws}}{4\pi^2 v_w g_* T_{n1}}\int_0^\infty dz \mu_{B_L}(z) \exp\left(-\frac{45\Gamma_{ws}}{4v_W}\right)\,,
\end{equation}
where $\Gamma_{ws}\simeq 1.0\cdot 10^{-6} T_{n1}$
is the weak sphaleron rate inside bubble \cite{sphaleron-ws}. Fig. \ref{figmu} shows the solutions to the transport equations for $\mu_i$ and $u_i$ for the BP1 and the BP2, which
give rise to the BAU, $Y_B\simeq 8.4\times 10^{-11}$ for the BP1 and $Y_B\simeq 8.3\times 10^{-11}$ for the BP2.

Noticed that the effective potential $V_{eff}$ have a $Z_2$ symmetry under which 
\beq
h_3\to -h_3,~~~~S\to -S.
\eeq
Therefore, there
will not be a bias between transitions to $(<h_1>,~ <h_2>,~ <h_3>,~ <S>)$ and $(<h_1>,~ <h_2>,~ -<h_3>, -<S>)$ from the origin (0,~0,~0,~0) GeV.
Thus, there are two
kinds of bubbles relating to $\theta_t$ and $-\theta_t$, which produce
baryon asymmetry of opposite signs. Eventually, the averaged baryon number is zero
in whole region due to their opposite signs. A soft $Z_2$ symmetry breaking term, $-\mu_3$$S^3$ can be introduced to 
solve the problem. For the BP2, the temperature of the  $Z_2$-breaking PT is significantly higher than $T_{n1}$
of the electroweak PT, and the regions with $-<S>$ can vanish when the electroweak
PT takes place. The needed condition is $\Delta V/T^4>10^{-16}$ with $\Delta V$ being the potential difference between
the vacua with $\pm<S>$ \cite{1110.2876,McDonald}. The $\mu_3$ with a value of $\order( 10^{-14})$ GeV can realize the condition for the BP2.
Unlike the BP2, the vacua with $\pm<S>$ for the BP1 are still around at the time of the electroweak PT, but the volumes occupied by the $\pm<S>$ phases
can be significantly different. One can approximately estimate the ratio between the number densities of bubbles with positive baryon number ($N_+$) and negative baryon number ($N_-$)
 \cite{9304267,9110342},
\beq
\frac{N_+}{N_-}={\rm exp}(-\Delta S_3/T),
\eeq
with $\Delta S_3$ being the $S_3$ difference between two types of bubbles. The global baryon density is given by
\beq
Y_B=Y_B^+\frac{N_+ - N_-}{N_+ + N_-},
\eeq
where $Y_B^+$ is the BAU generated from the bubble with $+<S>$. For the BP1, $Y_B \sim \frac{Y_B^+}{2}$ needs $\mu_3 \sim \order(10^{-1})$ GeV, and such value 
is incompatible with the expected spontaneous CP violation since the $\mu_3$$S^3$ term breaks the CP symmetry explicitly. 

\section{gravitational wave}
There are three sources of GW production at a first-order PT: bubble collisions, sound waves in the plasma and magnetohydrodynamic turbulence. 
We will focus on the GW spectrum from the sound waves in the plasma, which
typically is the largest contribution among them.
In addition to the two parameters $\beta$ and $\alpha$ describing the dynamics of the PT, the GW spectra depends on the wall velocity with respect to the plasma at infinite distance, $\tilde{v}_W$.
Note that $\tilde{v}_W$ can be significantly different from $v_W$ \cite{1103.2159}, which is the relative wall velocity
to plasma in front of the wall, and relevant for baryogenesis. We take $\tilde{v}_W=0.6$ in our calculation.

The GW spectrum from the sound waves can be expressed by \cite{gw-sw}
\begin{eqnarray}
\Omega_{\textrm{sw}}h^{2} & \ = \ &
2.65\times10^{-6}\left( \frac{H_{n}}{\beta}\right)\left(\frac{\kappa_{v} \alpha}{1+\alpha} \right)^{2}
\left( \frac{100}{g_{\ast}}\right)^{1/3} \tilde{v}_W\nonumber \\
&&\times  \left(\frac{f}{f_{sw}} \right)^{3} \left( \frac{7}{4+3(f/f_{\textrm{sw}})^{2}} \right) ^{7/2} \Upsilon(\tau_{sw})\ ,
\label{eq:soundwaves}
\end{eqnarray}
where $f_{\text{sw}}$ is the present peak frequency of the spectrum,
\begin{equation}
f_{\textrm{sw}} \ = \ 
1.9\times10^{-5}\frac{1}{\tilde{v}_W}\left(\frac{\beta}{H_{n}} \right) \left( \frac{T_{n}}{100\textrm{GeV}} \right) \left( \frac{g_{\ast}}{100}\right)^{1/6} \textrm{Hz} \,.
\label{fsw}
\end{equation}
The $\kappa_{v}$ is the fraction of latent heat transformed into the kinetic energy of the fluid \cite{1004.4187},
\beq
\kappa_v \simeq \kappa_B + (\tilde{v}_W-c_s) \delta\kappa +\frac{(\tilde{v}_W-c_s)^3}{(\xi_J-c_s)^3}\left[\kappa_C-\kappa_B-(\xi_J-c_s)\delta\kappa\right] ~({\rm for}~ c_s<\tilde{v}_W<\xi_J),
\eeq
with the sound velocity $c_s=\sqrt{1/3}$ and
\bea
&&\kappa_B \simeq \frac{\alpha^{2/5}}
{0.017 + (0.997 + \alpha)^{2/5}},~~~\kappa_C \simeq \frac{\sqrt{\alpha}}
{0.135 + \sqrt{0.98+\alpha}},\nonumber\\
&&\xi_J \simeq  \frac{\sqrt{\frac{2}{3}\alpha+\alpha^2}+\sqrt{1/3}}{1 + \alpha},~~~\delta \kappa\simeq -0.9\log\frac{\sqrt{\alpha}}{1+\sqrt{\alpha}}.
\eea
The suppression factor \cite{2007.08537}
\beq
\Upsilon(\tau_{sw})=1-\frac{1}{\sqrt{1+2\tau_{sw}H_n}},
\eeq 
arises due to the finite lifetime $\tau_{sw}$ of the sound waves \cite{2003.07360,2003.08892},
\beq
\tau_{sw}=\frac{\tilde{v}_W(8\pi)^{1/3}}{\beta \bar{U}_f}, ~~ \bar{U}^2_f=\frac{3}{4}\frac{\kappa_v\alpha}{1+\alpha}.
\eeq

\begin{figure}[tb]
 \epsfig{file=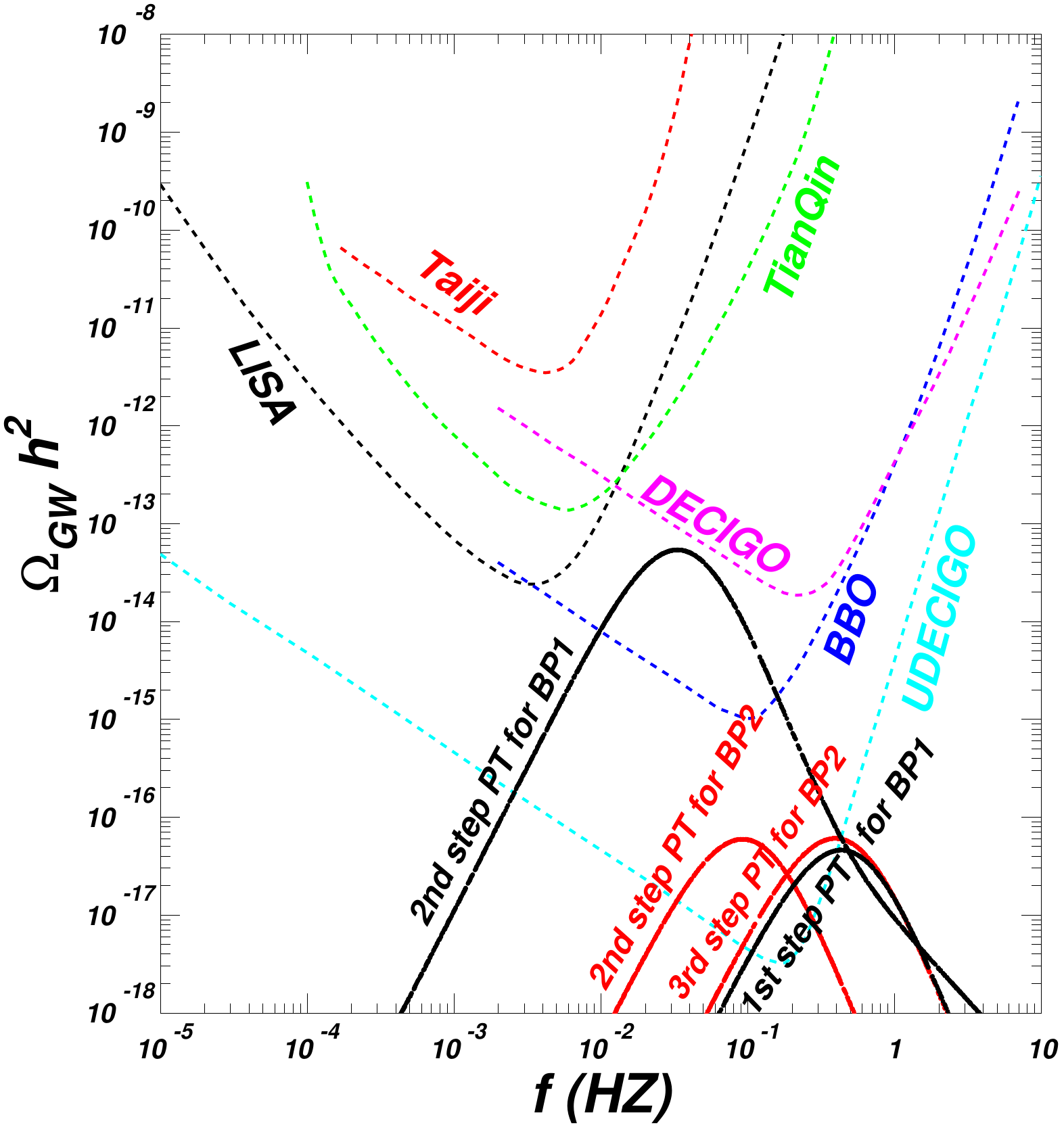,height=8cm}
\vspace{-0.4cm} \caption{Gravitational wave spectra for the BP1 and the BP2.} \label{figgrav}
\end{figure}

We examine the GW spectra for the BP1 and BP2, which are shown along with expected sensitivities of various future
interferometer experiments in Fig. \ref{figgrav}. 
For the BP1, the GW spectra from the first-step and second-step PTs have the peak frequencies around 0.44
Hz and 0.03  Hz, and the peak strengths of the former and the latter exceed the sensitivity curves of U-DECIGO and BBO, respectively.
For the BP2, the GW spectra from the second-step and third-step PTs have the peak frequencies around 0.09
Hz and 0.39  Hz, whose peak strengths exceed the sensitivity curves of U-DECIGO.
For the BP2, the superposed GW spectra from the second-step and third-step have an explicit double peaks, which can be observed by the U-DECIGO.
Note that there is still highly uncertainty on the value of $\Upsilon(\tau_{sw})$,
and its determination
needs considerable numerical simulations and analytical
insights in the future. In addition, a full exploration of
the parameter space will potentially find more promising regions for detectable two-peaked GW signal at the U-DECIGO.

\section{Conclusion}
In a singlet pseudoscalar extension of 2HDM, we studied the spontaneous CP violation EWBG via two-step PTs and three-step PTs, 
and took the BP1 and the BP2 to perform detailed calculations.
The first-step of the two-step PTs is a SFOEWPT, which converts ($<h_1>$, $<h_2>$, $<h_3>$, $<S>$) into an electroweak symmetry broken phase from (0, 0, 0, 0) GeV, and
breaks the CP symmetry spontaneously. The electroweak sphaleron processes bias the CP asymmetry into the baryon number during the first-step PT.
Also the second-step of the two-step PTs is a SFOEWPT, which converts the phase into the observed vacuum at zero temperature, and
the CP-symmetry is restored. 
However, the vacua with $\pm<S>$ are still around at the time of the first SFOEWPT, and an explicit CP-violation term, $-\mu_3 S^3$ with $\mu_3\sim (10^{-1})$ GeV, is required to
guarantee the volumes occupied by the $\pm<S>$ phases to be significantly different, leading to a sufficient baryon number density.
The first-step of the three-step PTs is a second-order PT during which the $S$ field firstly develops a nonzero VEV, and VEVs of $h_1$, $h_2$, and $h_3$ still remain zero.
Similar to the case of two-step PTs, the observed BAU is produced via the EWBG mechanism at the second-step.
The third-step of the three-step PTs is a SFOEWPT, which converts the phase into the observed vacuum at the zero temperature and restores the CP symmetry.
A very tiny CP-violation term, $-\mu_3 S^3$ with $\mu_3\sim (10^{-14})$ GeV, is required to guarantee the regions with $-<S>$ to disappear when the second-step PT takes place. 
 Meanwhile, the GW spectra through the two-step and three-step PTs can reach the sensitivities of BBO and U-DECIGO. Even more interesting is that a
 two-peaked GW signal could be observed at the U-DECIGO.

\section*{Acknowledgment}
We thank Wei Chao, James M. Cline, and Yang Zhang for helpful discussions. We are grateful to
Wei Chao for reading the manuscript. This work was supported by the National Natural Science Foundation
of China under grant 11975013.

\end{document}